\documentclass[a4paper]{article}

\usepackage{INTERSPEECH2022}
\usepackage{multirow}
\usepackage{tipa}
\usepackage{subcaption}
\usepackage{hyperref}
\usepackage{IEEEtrantools}

\title{Speaker Anonymization with Phonetic Intermediate Representations}

\name{Sarina Meyer, Florian Lux, Pavel Denisov, Julia Koch, Pascal Tilli, Ngoc Thang Vu}
\address{Institute for Natural Language Processing (IMS), University of Stuttgart, Germany}
\email{sarina.meyer@ims.uni-stuttgart.de}

\begin{document}
\bstctlcite{IEEEexample:BSTcontrol}

\maketitle
\begin{abstract}
In this work, we propose a speaker anonymization pipeline that leverages high quality automatic speech recognition and synthesis systems to generate speech conditioned on phonetic transcriptions and anonymized speaker embeddings. 
Using phones as the intermediate representation ensures near complete elimination of speaker identity information from the input while preserving the original phonetic content as much as possible.
Our experimental results on LibriSpeech and VCTK corpora reveal two key findings: 
1) although automatic speech recognition produces imperfect transcriptions, our neural speech synthesis system can handle such errors, making our system feasible and robust, and 
2) combining speaker embeddings from different resources is beneficial and their appropriate normalization is crucial. 
Overall, our final best system outperforms significantly the baselines provided in the Voice Privacy Challenge 2020 in terms of privacy robustness against a lazy-informed attacker while maintaining high intelligibility and naturalness of the anonymized speech. 
\end{abstract}
                                                                                                
\section{Introduction}
With the advances in machine learning, speech interfaces have become increasingly popular in various applications. A prominent example are virtual assistants, which appear to be available in almost every day situations - ranging from assistants at home, in smart phones to assistants in cars. 
Such applications send recorded speech signals to centralized servers where the signals get processed and potentially stored. 
Considering that speech data carries information about the individuals' identity, it is defined as personal data.
Thus, with new regulations such as the General Data Protection Regulation (GDPR) in the European Union, which strengthen privacy preservation and protection of personal data \cite{eureg}, the demand and importance of voice privacy methods rise.
Consequently, a Voice Privacy Challenge (VPC) was launched in 2020 \cite{tomashenko2020introducing}.
This challenge serves as a starting point to enable research on this topic. 
The objective is to outsmart speaker verification systems by removing the speaker's identity from the speech signal as much as possible while still preserving the linguistic content.

The evaluation of the VPC \cite{tomashenko2022voiceprivacy} showed that most speaker anonymization approaches follow one of the baselines and can thus be divided into two groups, with the majority focusing on the first one: systems based on (1) x-vector \cite{xvect} embeddings, and on (2) signal processing techniques. 
Main findings in the challenge include a degradation in naturalness and intelligibility of the speech by all methods, although in less extent when using signal processing, but a better performance of the x-vector based approaches according to objective privacy and speech recognition metrics. However, speaker identity information has been found to be encoded not only in the x-vectors but also in pitch and speech recognition bottleneck features which are used for resynthesis in an unaltered form and thus decrease the anonymization strength of the approaches. Since most systems of group (1)  focus only on the x-vector modification by leaving the other modules unchanged, this problem is rarely addressed. 
Exceptions are \cite{champion2020speaker} with a semi-adversarial approach to mask speaker information in linguistic content, \cite{champion2021astudy} by modifying pitch before synthesis, and \cite{mawalim2022speaker} by modifying pitch and duration of the input speech prior to feature extraction. However, these approaches lead only to small or no gain in privacy as compared to the baselines. To our knowledge, no approach based on x-vectors has yet aimed to suppress privacy leakage without relying on pitch or bottleneck features for synthesis.

We hypothesize that removing all speech information except for the transcription will eliminate speaker identity footprints encoded in speech and therefore allows a fully anonymized setting. 
In order to verify this hypothesis, we propose a speaker anonymization system that combines high quality (i.e., state-of-the-art-level) automatic speech recognition  (ASR) and text-to-speech (TTS) systems to convert speech to phones and thus to eliminate speaker identity information completely before synthesizing anonymized speech conditioned on modified speaker embeddings. 
As this introduces unexplored design questions and potential points of error propagation, we tackle 
the following research questions: (1)  Is it better to use text or phones as intermediate representation and input to the TTS module?, (2) How do ASR errors impact the synthesized speech?, and (3) Which type of speaker embeddings is most suitable in this setting?

The performances of the pipeline and each design choice are evaluated using the framework of the Voice Privacy Challenge. We observe that reducing the linguistic content to phones increases the privacy of the anonymized speech and is robust against speaker verification attacks even if the attacker uses anonymized enrollment data. Furthermore, we find that combining two embedding techniques, x-vector and ECAPA-TDNN \cite{DBLP:conf/interspeech/DesplanquesTD20}, results in the best synthesized speech, and we see a high importance of embedding normalization before synthesis in order to generate distinctive voices. Finally, we experience the TTS system directly using the recognized phone sequences instead of text to be inherently robust against ASR errors. 
The code and a test demo of the system are available online\footnote{\href{https://github.com/DigitalPhonetics/speaker-anonymization}{https://github.com/DigitalPhonetics/speaker-anonymization}}.

\section{Speaker Anonymization System} \label{sec:system}
\subsection{Overview}
\begin{figure}[h]
    \centering
    \includegraphics[width=\linewidth]{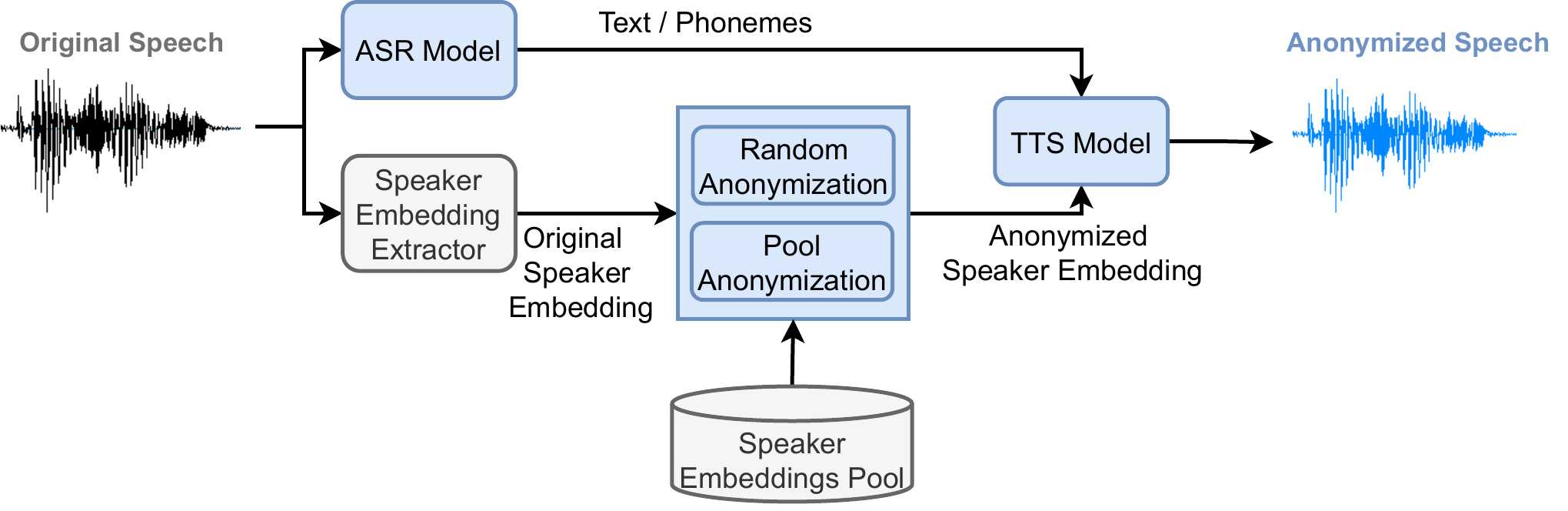}
    \caption{Our proposed anonymization system} 
    \label{fig:system}
\end{figure}

Figure \ref{fig:system} displays the anonymization pipeline.
Our system consists of three major parts: (1) an ASR system, (2) a speaker embedding anonymizer, and (3) a TTS model.
A speech signal is anonymized by feeding its recognized transcription in form of word or phone sequences as well as a modified version of a vector representing the speaker identity into the TTS system to resynthesize the signal.

\subsection{Speaker embedding modification}
\label{sec:spk_anon}

Contrary to previous work, we do not use x-vectors \cite{xvect} as our only source of speaker identity information but combine them with the recent ECAPA-TDNN embeddings \cite{DBLP:conf/interspeech/DesplanquesTD20}. 
Each 704-dimensional vector consists of the speaker's ECAPA embeddings in the first 192 positions and of its x-vector in the remaining 512 positions. \footnote{Preliminary experiments show that their concatenation makes the speaker vector space more separable (see Section \ref{subsubsec:impact_embeddings}), suggesting that both methods contain information that is complementary to each other.}

An analysis of these embeddings revealed that each dimension contains a different value distribution such that the range of values varies a lot between the dimensions. For instance, the position for which the smallest value range could be found in the original speaker embeddings, the values range between -0.97 and -0.11, while the dimension with the largest range allows values between -72.37 and 81.59. Ignoring these differences during the anonymization could lead to unnatural embeddings with properties that would not be found for a real speaker.

\subsubsection{Random anonymization}
A random modification of each speaker embedding serves as a simple baseline. For each position in the speaker vector, we randomly choose a value in the corresponding valid value range. This random baseline is simple and independent of training processes but might produce embeddings that are unrealistic or too similar to the original speaker embeddings.

\subsubsection{Pool anonymization} \label{subsubsec:anon_vpc}
The second approach is based on the primary baseline of the VPC \cite{tomashenko2020introducing}. It constructs a pool vector space extracted from speakers of the training set, and trains a PLDA model as a distance measure between the vectors. During anonymization, the 200 pool speaker embeddings that are most distant from the given speaker are selected, and a new speaker embedding is computed as the average between a random subset of 100 vectors of them. This new embedding serves as the anonymized speaker embedding. To ensure the correct value ranges, after selecting an anonymized vector, we scale each dimension separately to lie within the range found in the original embedding space.

\subsection{Speech to speech}
\subsubsection{ASR}

Our end-to-end ASR system follows hybrid CTC/attention architecture \cite{watanabe2017hybrid} and relies on Conformer encoder \cite{gulati2020conformer} and Transformer decoder. Its implementation is derived from the LibriSpeech recipe of the ESPnet2 toolkit \cite{watanabe20212020}. SpecAugment \cite{park2019specaugment} and 3-way speed perturbation \cite{ko2015audio} data augmentation methods are utilized during the training. We use SentencePiece \cite{kudo2018sentencepiece} to learn unigram language model \cite{kudo2018subword} subword units from the training data transcriptions in speech-to-text (STT) and text-to-speech (TTS) formats as well as from phonemized TTS transcriptions. Respectively, we train and evaluate three ASR systems differing by the type of output units.

\subsubsection{TTS}
We train a FastSpeech 2 \cite{ren2020fastspeech} speech synthesis system that uses the Conformer architecture \cite{gulati2020conformer} in both encoder and decoder with the IMS Toucan toolkit \cite{lux2021toucan} which is in turn based on the ESPnet toolkit \cite{hayashi2020espnet, hayashi2021espnet2}. The inputs to the system are articulatory feature vectors rather than the embedding-lookup table used in most TTS systems, as introduced in \cite{lux2022LAML}.
Furthermore we apply FastPitch's \cite{lancucki2021fastpitch} control mechanism to FastSpeech 2's \cite{ren2020fastspeech} energy and pitch predictions.
To perform the spectrogram inversion needed to convert the synthesized spectrograms into waveforms, we make use of the HiFi-GAN architecture \cite{kong2020hifi} as implemented in the IMS Toucan toolkit \cite{lux2021toucan}.
In order to handle the zero-shot speaker adaptation required for the generation of speech with synthetic embeddings, we condition the TTS on the ensemble of speaker embeddings described in Section \ref{sec:spk_anon} by concatenating them to the encoder output and then projecting them back into the dimensionality of the encoder output.

\subsubsection{Connecting ASR and TTS}
The range of models we explore include ASR models which predict texts that are then phonemized as well as ASR models that directly predict phones. Although ASR and TTS are trained separately, we investigate TTS models which are finetuned on ASR outputs in order to mitigate systematic ASR errors. Mitigating ASR errors may however already happen implicitly because we map the phones that the ASR recognizes to articulatory vectors, which are then fed into the TTS system. Unlike a purely identity based representation of the phones, the articulatory vectors relate phones to each other.

\section{Experimental Setup} \label{sec:exp_setup}
\subsection{Data}
For the experiments, we use the same datasets as employed in the VPC. That is, the speaker embeddings have been trained by SpeechBrain \cite{speechbrain} on VoxCeleb 1 and 2 \cite{Nagrani19, Nagrani17, Chung18b}. The speech synthesis data LibriTTS-600 \cite{zen2019libritts} consisting of LibriTTS-clean-100 and LibriTTS-other-500 is used for training the TTS module. Two versions of the ASR model were created, one on the LibriSpeech-600 corpus \cite{panayotov2015librispeech}, the speech recognition counterpart of LibriTTS-600, and the other one directly trained on LibriTTS-600 to generate output that is more suitable for the TTS module. All training for the anonymization models is performed on the LibriTTS-other-500. 
The evaluation of the system is performed on the development and test data provided by the VPC which are described in \cite{tomashenko2020introducing}.

\subsection{Evaluation metrics}
The quality of the anonymization is assessed with the automatic speaker verification (ASV) model provided by the VPC 2020. It measures the performance using the Equal Error Rate EER and the discrimination loss $C_{llr}^{min}$ as described in \cite{tomashenko2020introducing, tomashenko2022voiceprivacy}.
We aim for a score close to 50\% denoting random prediction. 
The ASV tests are conducted in two attack scenarios, 
in which either (a) only the trial data is anonymized (\textbf{O-A}, for \textbf{o}riginal enrollment and \textbf{a}nonymized trial data), or (b) in which both enrollment and trial data are anonymized but by using different target speakers for each set (\textbf{A-A}, also called lazy-informed).  Anonymization should not only lead to the speaker be un-identifiable but also to retaining the distinctiveness of different speakers. 
Thus, this distinctiveness is measured using the De-Identification (DeID, [0, 1]) and Gain of Voice Distinctiveness (GVD, [$-\infty$, $\infty$]) metrics proposed in \cite{noe20speech}. Most desirable are DeID scores close to 1 (high de-identification), and GVD ones close to 0 (same voice distinctiveness as in original data) or higher (increased voice distinctiveness).
Finally, word error rates (WER) are also calculated for the original and anonymized speech recordings.

\subsection{Baseline}
As we use the same data and evaluation strategies as in the VPC 2020, we can easily compare our approaches to its primary baseline\footnote{Since the challenge uses several different evaluation metrics and could not identify a clear winning system for all of them, a comparison to the best approaches is not straight-forward. We refer interested readers to the results paper of the challenge \cite{tomashenko2022voiceprivacy} that can be used for a metric-by-metric comparison between all participants of the challenge and our approach.}. It uses the same anonymization technique as in our pool approach but with x-vector embeddings and without normalization, and is embedded in a different ASR-TTS pipeline based on pitch and bottleneck features.

\section{Experimental Results} \label{sec:exp_results}
In order to evaluate different settings for each of the three components of the system
, we keep in each experiment two components fixed and alter only the third one. If not stated otherwise, we (a) anonymize the speaker embeddings using the speaker pool technique, (b) use directly the phone sequence as given by the ASR model without transforming it into text, and (c) synthesize the speech by the TTS system that is trained only on the clean Libri-100 data without any finetuning.

\subsection{Anonymization}
\begin{table*}[th]
    \footnotesize
    \caption{Privacy results for different anonymization strategies. All systems except for the baseline (BL) use the phone output of the ASR systems and the TTS trained on Libri-100. Pool raw stands for the pool anonymization without normalization of embeddings.}
    \label{tab:anon_strategies}
    \centering
    \begin{tabular}{cc|cc|cc|cc|cc|cc|cc}
        \toprule
        &  & \multicolumn{6}{|c}{Female} & \multicolumn{6}{|c}{Male} \\
          &  & \multicolumn{2}{c|}{O - A} & \multicolumn{2}{c|}{A - A}  & & &  \multicolumn{2}{c|}{O - A} & \multicolumn{2}{c|}{A - A} & & \\
        Data & Anon & EER & $C_{llr}^{min}$ & EER  & $C_{llr}^{min}$ & DeID & GVD &  EER  & $C_{llr}^{min}$ & EER & $C_{llr}^{min}$ & DeID & GVD\\
        \midrule
        \multirow{4}{*}{Libri} & \textit{BL} & 47.26 & 1.00 & 32.12 & 0.84 & 0.98 & -10.09 & \textbf{52.12} & 1.00 & 36.75 & 0.90 & 1.00 & -8.95 \\
         & pool & 56.75 & 0.94 & 51.64 & 0.96 & 0.98 & -0.14 & 45.66 & 0.96 & 43.88 & 0.95 & 1.00 & \textbf{-0.11} \\
        &  random & 53.47 & 1.00 & \textbf{50.91} & 1.00 & 1.00 & \textbf{-0.13} & 42.32 & 0.96 & 47.22 & 0.96 & 0.98 & -0.14\\
        &  pool raw & \textbf{50.00} & 0.98 & 51.64 & 1.00 & 1.00 & -6.50 & 44.54 & 0.98 & \textbf{52.56} & 1.00 & 1.00 & -7.68\\
        \midrule
        \multirow{4}{*}{VCTK} & \textit{BL} & 48.05 & 1.00 & 31.74 & 0.85 & 1.00 & -10.56 & 53.85 & 1.00 & 30.94 & 0.83 & 1.00 & -11.58 \\
         & pool & \textbf{51.34} & 0.99 & 48.46 & 0.99 & 1.00 & \textbf{-0.02} & \textbf{50.75} & 1.00 & 40.59 & 0.96 & 1.00 & \textbf{-0.58} \\
        & random & 52.98 & 1.00 & 55.40 & 1.00 & 1.00 & -0.88 & 52.76 & 1.00 & \textbf{45.12} & 0.96 & 1.00 & -1.38 \\
        & pool raw & 54.22 & 1.00 & \textbf{49.54} & 1.00 & 1.00 & -7.79 & 53.04 & 1.00 & 43.23 & 0.98 & 1.00 & -9.57\\
        \bottomrule
    \end{tabular}
\end{table*}

The speaker verifiability results for different anonymization strategies and scenarios are given in Table \ref{tab:anon_strategies}. While the proposed methods show a similar anonymization strength as the baseline if the attacker uses the original data for enrollment (EER $\approx$ 50), they clearly outperform it in the fully anonymized A-A scenario (EER $\approx$ 30 vs. EER $\approx$ 50). This shows that our approach is indeed more successful in concealing identity information in all modules, not just the speaker embeddings. The findings from the ASV attacker are confirmed by the de-identification scores. Of special interest is the performance in terms of gain of voice distinctiveness since the baseline and all similar approaches in the challenge produced voices that lacked the original distinctiveness of the data. This drawback is eliminated in our system but only if the anonymized speaker embeddings are normalized accordingly (compare \textit{pool} and \textit{pool raw}). This observation is further addressed in Section \ref{subsubsec:impact_normalization}.

\subsection{Utility}
\begin{table}[h]
    \caption{ASR results as WER for the system with different settings in the ASR, TTS and anonymization modules.}
    \label{tab:wer_results}
    \centering
    \footnotesize
    \begin{tabular}{cc|cc}
        \toprule
        ASR & TTS  & Libri & VCTK \\
        \midrule
        \multicolumn{2}{c|}{\textit{No anonymization}} & 4.15 & 12.82\\
        \multicolumn{2}{c|}{\textit{Baseline}} & \textbf{6.73} & 15.23\\
        \midrule
        STT & Libri100  & 9.89 & 18.89\\
        TTS & Libri100  & 9.42 & 17.53\\
        phones & Libri100 & 7.55 & \textbf{13.83}\\
        phones & Libri100 + finetuned & 8.38 & 16.33\\
        phones & Libri600 & 8.87 & 14.96\\
        phones & Libri600 + finetuned & 9.97 & 16.12\\
        \bottomrule
    \end{tabular}
\end{table}

Table \ref{tab:wer_results} presents the utility evaluation results in terms of WER values obtained after the decoding of the original and anonymized speech using the ASR system from the VPC.  Phonetic output of ASR clearly has an advantage over STT or TTS style output units, bringing the utility of our system closer to the baseline on LibriSpeech test set and exceeding it on VCTK test set.
WER values are higher for the TTS systems trained on ASR predictions or on the larger dataset or both, what is consistent with the privacy metrics and indicates that the noisier training data makes the TTS training less stable.

\section{Analysis} \label{sec:analysis}
\subsection{Quantitative analysis}
\subsubsection{Impact of speaker embedding normalization} \label{subsubsec:impact_normalization}
One problem that arises from using the pool anonymization is that the value range of the speaker embeddings change with the anonymization. As can be expected, the TTS works best if the speaker embeddings follow the same information structure as they do during training. The difference becomes quite clear in Table \ref{tab:anon_strategies}. The GVD of the pool raw approach (unscaled embeddings) shows that the system is basically unusable since it produces a very similar voice regardless of the anonymized speaker embedding fed to the TTS. Since pool anonymization works by averaging speaker embeddings and the value range of the dimensions is in most cases centered around 0, the range of values in the anonymized embeddings becomes smaller, leading to the collapse of different speakers into one. By simply scaling each of the dimensions in the anonymized speaker embeddings according to their usual range of values, we achieve near perfect GVD using the pool approach.

\subsubsection{Impact of ASR type}
ASR and TTS systems are typically trained on transcriptions of different types that we denote as TTS and STT. We considered to train the ASR system on TTS transcriptions in order to avoid this mismatch and to eliminate errors potentially caused by it. One step further was to convert TTS formatted transcriptions to phones and to train ASR to predict phones in order to get even closer to what TTS system receives as its input. 
Our results show that type of ASR output units does not have prominent influence on the anonymization results of the system. However, as mentioned earlier, the utility properties of system output are noticeably better with the phone based ASR. Furthermore, phonetic intermediate representation removes an extra step of phonetic conversion from the inference pipeline and makes the whole anonymization system slightly simpler and faster.

\subsubsection{Impact of error propagation}
Because of the sequential design of our system, errors that the ASR makes at the very beginning propagate throughout the entire pipeline, affecting primarily the WER of the anonymized speech (see Table \ref{tab:wer_results}). We experimented on two ways of mitigating this effect by adapting the TTS:  
1) Using more data should in theory lead to a more robust TTS. We see however that when we include the other-500 training data of the LibriTTS corpus, the overall performance with respect to WER decreases. This is likely due to the noisy data in the other-500 set which causes the TTS to become less intelligible in general. 
2) Finetuning the TTS on ASR outputs in order to learn how to deal with systematic errors that the ASR makes also shows no improvements. We get the best performance when training the TTS on only the gold transcripts of the clean-100 subset of LibriTTS.
While it is counter-intuitive that neither more data nor finetuning the joint between the models bring any benefits, we suspect that the reason lies in the TTS learning an implicit language model which is aware of how phone realizations change in context. This works best on the clean gold transcriptions and can improve cases where unnatural transitions are given as input. Furthermore, since the articulatory feature representations we use as the input representation for TTS directly relate phonemic units with each other, slight mistakes that the ASR makes, e.g. confusing [\textipa{a}] with [\textipa{\ae}] or [\textipa{\textturna}] are much less severe downstream.

\subsection{Qualitative analysis}
\subsubsection{Impact of speaker embeddings}\label{subsubsec:impact_embeddings}

The original speaker embeddings used in the system are formed by a concatenation of ECAPA and x-vectors. 
The decision for this method is based on an intrinsic evaluation of the clustering behavior of each embedding space and the subjective performance of the speech synthesis conditioned on the different vector types. Regarding the first assessment, we extracted the x-vectors and ECAPA embeddings for each utterance in the LibriTTS-other-500 data, and performed K-Means clustering on each vector space as well as the combined one created through concatenation. 
The combined embedding space resulted in the best clustering, showing that both vector types complement each other in terms of speaker distinctive information. By manually analyzing the synthesized speech for each embedding method, we found that using just one of the embedding functions worked well for multispeaker TTS with no unseen speakers. 
Combining the embeddings however significantly increased the variance in samples generated from random speaker embeddings. 
This benefits the zero-shot voice adaptation needed for dealing with speaker embeddings generated by the means described in Section \ref{sec:spk_anon}.

\subsubsection{Visualizing speaker embeddings}
\begin{figure}[h!]
     \centering
        \begin{subfigure}[b]{0.47\linewidth}
            \centering
            \includegraphics[width=\textwidth]{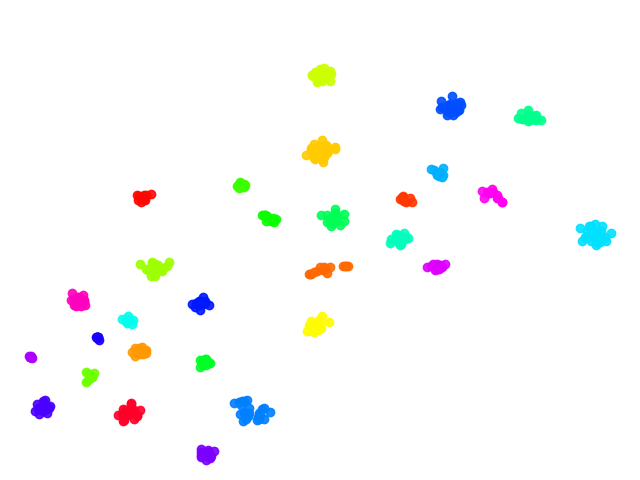}
            \subcaption{Original}
            \label{fig:libri_origina}
        \end{subfigure}
        \hfill
        \begin{subfigure}[b]{0.47\linewidth}   
            \centering 
            \includegraphics[width=\textwidth]{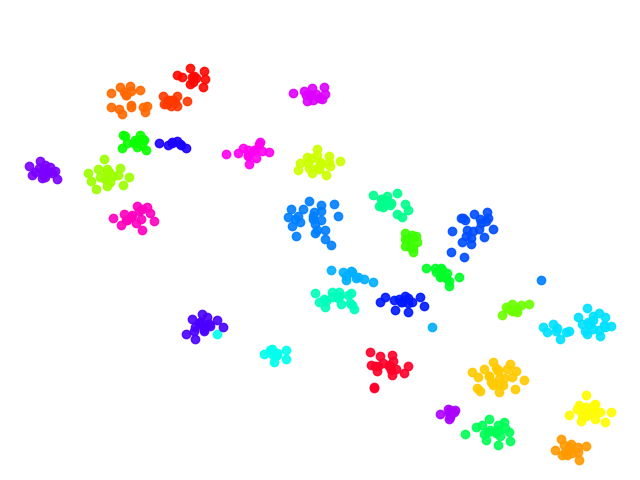}
            \subcaption{Anonymized}%
            \label{fig:libri_anon}
        \end{subfigure}
        \caption{Speaker embeddings before and after normalized pool anonymization and synthesis of the LibriSpeech enrollment data. Each color denotes one speaker. Projection to 2D-space is done with t-SNE \cite{van2008visualizing}.}
        \label{fig:spk_emb}
\end{figure}

We present a visualization of original and anonymized speaker embeddings for a subset of LibriSpeech in Figure \ref{fig:spk_emb}. 
We observe that, although not as dense as with the original embeddings, the clusters computed from anonymized speech are clearly distinguishable for the most part. This does not only confirm that the system produces embeddings that are distinctive between speakers while being consistent within a speaker; it also shows that our TTS is able to generate speech from the anonymized embeddings with the same properties. Furthermore, the plot shows that speakers which are close to each other in the original space are not necessarily close after anonymization. We take this as further evidence for successful anonymization.

\section{Conclusions} \label{sec:conclusion}
This study presents a speaker anonymization system that minimizes the identity information by resynthesizing speech based only on the phonetic content representation and an anonymized speaker embedding. Using a pipeline containing high quality speech recognition and synthesis components and simple embedding modifications, we obtain successfully anonymized speech while retaining linguistic information. The system outperforms previous work in terms of privacy robustness against lazy-informed attacks. In future work, we plan to enhance the system by a more sophisticated anonymization component which learns how natural speaker embeddings are structured and how to imitate this in order to create artificial embeddings, for instance in an adversarial training setting. We believe naturalness to be a crucial condition for anonymized speaker embeddings for leading to natural-sounding anonymized speech. 

\bibliographystyle{IEEEtran}

\bibliography{mybib}

\end{document}